# A room temperature single-photon source based on strongly interacting Rydberg atoms

F. Ripka[1], H. Kübler[1], R. Löw[1], and T. Pfau[1]*

[1] 5. Physikalisches Institut, Universität Stuttgart, Center for Integrated Quantum Science and Technology, Pfaffenwaldring 57, 70569 Stuttgart, Germany

*To whom correspondence should be addressed. E-mail: t.pfau@physik.uni-stuttgart.de
(Date: June 6, 2018)

Tailored quantum states of light can be created via a transfer of collective quantum states of matter to light modes. Such collective quantum states emerge in interacting many-body systems if thermal fluctuations are overcome by sufficient interaction strengths. Therefore, typically ultracold temperatures or strong confinement are required. We show that the exaggerated interactions between giant Rydberg atoms allow for collective quantum states even above room temperature. The emerging Rydberg blockade allows then only for a single Rydberg excitation. We experimentally implement a four-wave mixing scheme to demonstrate an on-demand single-photon source. The combination of glass cell technology, identical atoms, and operation around room temperature promises scalability and integrability. This approach has the potential for various applications in quantum information processing and communication.

Single-photon emitters are of interest for manifold applications in quantum computation *(1)*, simulation *(2)*, sensing *(3)*, and especially in quantum secure communication *(4)*. In particular, the latter will demand efficient quantum repeaters *(5)* that require efficient single-photon creation and storage schemes, ideally at the same platform. On the one hand, room temperature alkali gases can perfectly serve as such storage media and second-long storage times have been already demonstrated for weak coherent states *(6)*. On the other hand, the generation of single photons with stable wavelength and adapted bandwidth is still a challenge.

The first observation of anti-bunched photon counting statistics from single-atom fluorescence dates back to 40 years ago *(7)*. Today very efficient on-demand solid state sources based on quantum dots *(8)*, color centers *(9)*, and single molecules *(10)* are available. Although in principle most of these systems are working at room temperature as well *(11 - 13)*, they do require cryogenic temperatures for optimal performance. As they are solid-state embedded, they suffer from interactions with phonons, spin noise, strain, and drifting electric fields. These result in large variations in frequencies, spectral wandering, and additional phase noise causing spectral diffusion, which is a fundamental limitation for scalability.

In contrast, atoms *(7)* and ions *(14)* inherently produce spectrally indistinguishable single photons. Here, the highest fidelities have been achieved with ultra-cold atoms inside high-finesse cavities *(15)*. Even first small photonic networks have been realized with them *(16)*. With room temperature gases, so far, heralded single photons with sub-Doppler linewidth can be created *(17)*. However, like parametric down-conversion *(18)*, they undergo a spontaneous generation process, and this requires a triggered memory for temporal synchronization of the created photons.

A rather new approach makes use of strongly interacting Rydberg atoms, which suppress multiple excitations within a certain radius *(19)*. This interaction mechanism can be converted into an effective optical non-linearity *(20)*, even on the single-photon level *(21)*. Based on the Rydberg blockade effect in ultra-cold samples, researchers have demonstrated photon anti-bunching *(22, 23)*, single-photon transistors *(24, 25)*, and even higher-order correlated light states *(26)*.

However, not only at ultra-low temperatures but also in thermal vapors, Rydberg-excited atoms can be coherently excited and optically probed *(27)*. Nanosecond-pulsed and intense excitation fields at GHz Rabi coupling strengths allow for coherent dynamics faster than one nanosecond per Rabi cycle *(28)*. On this time scale, thermal atoms move less than the optical wavelength and can be assumed nearly frozen. As soon as they are excited, the collective excitation can be retrieved within 1.2 nanoseconds, before dephasing effects set in *(29)*. To observe strong Rydberg interaction effects, the atom-atom interaction strength has to be larger than the excitation bandwidth. The equivalent of 3 GHz of van der Waals-type Rydberg interactions can be achieved in a dephasing regime where the excitation volume is larger than the interaction distance *(30)*.

Blockade of Rydberg excitations occurs at distances shorter than the blockade radius $r_B = (C_6/\hbar \Omega_{eff})^{1/6}$, with the van der Waals coefficient $C_6$ and the effective two-photon coupling strength *(31)* $\Omega_{eff}$. Let us assume the addressed Rydberg state be $40S_{1/2}$ and $\Omega_{eff} = 400$ MHz, then *(32)* $C_6 = 2\pi\hbar$ x 806.3 MHz $\mu m^6$ and the blockade radius results in $r_B = 1.1(1)$ µm. In order to suppress multiple excitations in an atomic ensemble by Rydberg blockade, the size of the ensemble has to be smaller than that radius. Rubidium atoms can be coherently excited to Rydberg states in vapor cells of such a size, without being limited by atom-wall interactions *(33)*. Therefore, it has been predicted that a four-wave mixing experiment in a microscopic cell will result in a suppression of multiple excitations and hence to an emission of single photons *(34)*. It is important to note that the size of such a small ensemble is still larger than the optical wavelength. This guarantees a directed emission pattern during the four-wave mixing process *(35)*. Besides, the emitted photon reflects the shapes of the excitation pulses and is not created in a spontaneous emission process. Furthermore, the emitted photons do not undergo significant jitter, neither in the center frequency nor in the bandwidth.

To truncate the Rydberg-excited ensembles to µm³ volumes, we focus the 795 nm excitation beam to a Gaussian $1/e^2$-waist of $w_0 = 1.45(5)$ µm (figure 1C). The vapor cell is designed in a wedge-shaped geometry and provides a distance range from 0 to several tens of micrometers (figure 1A and B). Thus, we can choose the degree of longitudinal confinement of the atomic ensemble. The cell is filled with rubidium at natural abundance and the experiment is done using the $^{85}$Rb isotope. The vapor pressure is set by a temperature of 140 °C, which corresponds to a steady-state density of approximately 30 atoms per µm³ in the hyperfine ground state F = 3. Yet, this is not sufficient for an efficient single-photon generation. However, as there is a large number of atoms attached to the surfaces of the glass windows, we desorb a fraction of them via light-induced atomic desorption *(36)*. A 2.5 ns pulse at 480 nm raises the number of atoms to more than 1,000 per µm³ within 5 ns (figure 2B). The Rydberg-excited volume contains then about 2,000 atoms. In the following few nanoseconds, the optical thickness decreases again as they hit the opposite surface.

The photons at 780 nm are generated in a pulsed four-wave mixing cycle about 5 ns after the pulsed desorption (figure 2A and B). The durations of the excitation pulses are 2.5 ns each. In the limit of adiabatic elimination of the intermediate $5P_{1/2}$ state, the Rabi frequency of the coupling between the ground and the Rydberg state is $\Omega_{eff}/2\pi = 0.41(16)$ GHz. The Rabi frequency of the transition from the Rydberg state to the excited state $5P_{3/2}$ is $\Omega_{480}/2\pi = 1.15(9)$ GHz. The photons at 780 nm are emitted into the same spatial mode as the 795 nm beam. They are collected by an aspheric high numeric aperture lens behind the cell. For spatial and spectral filtering, the emission is coupled into a single-mode fiber and frequency-filtered by bandpass filters and a temperature-stabilized etalon with a bandwidth of 1.7 GHz. Afterwards, the photons are detected in a Hanbury Brown-Twiss-type setup. The excitation pulses are recorded in order to enable a post-selection of events with respect to intensities and timing jitters (see Supplementary Material).

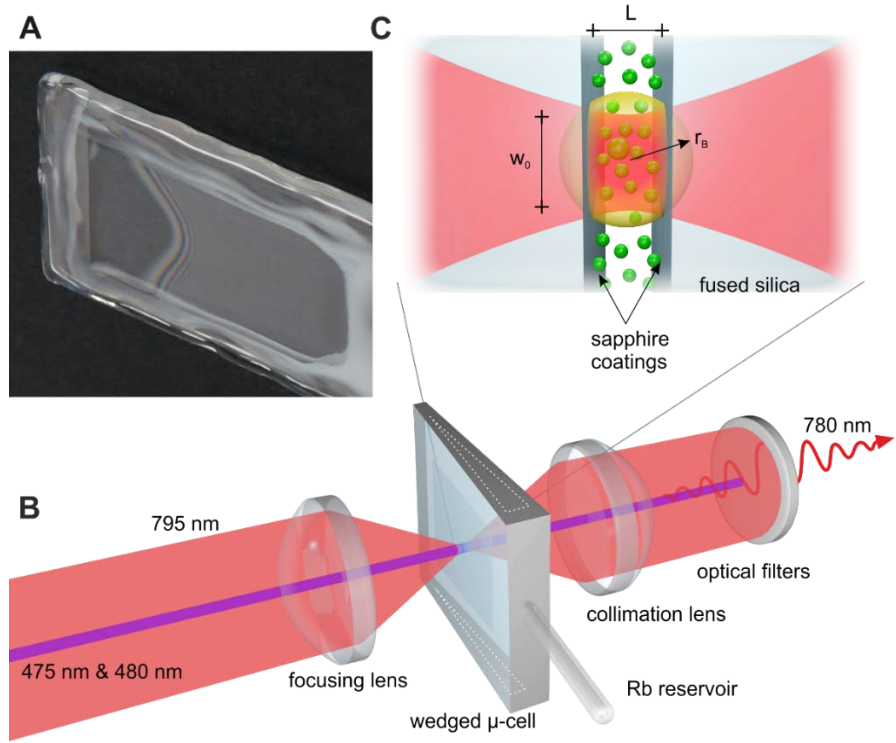

**Fig. 1. (A)** Atomic vapor cell and focus. Photograph of the of wedge-shaped micro-cell. The outer dimensions are 25 mm x 50 mm. The glass windows have a thickness of 1.21(1) mm, matching the high numeric aperture aspheric lenses (focusing and collimation), so that their diffraction limits are reached. The inner distance of the glass windows varies from touching each other to about 100 µm. The thicknesses are determined via residual-finesse resonances. In the region of 400 nm to few µm interference effects can be observed (so-called Newton's rings). **(B)** Schematic of the setup and, **(C)** zoom of excitation volume inside the vapor cell. The excitation of the rubidium atoms is transversally limited to the diameter of the 795 nm beam; 475 nm and 480 nm beams are much larger; all optical beams are co-propagating. More details in Supplementary Material.

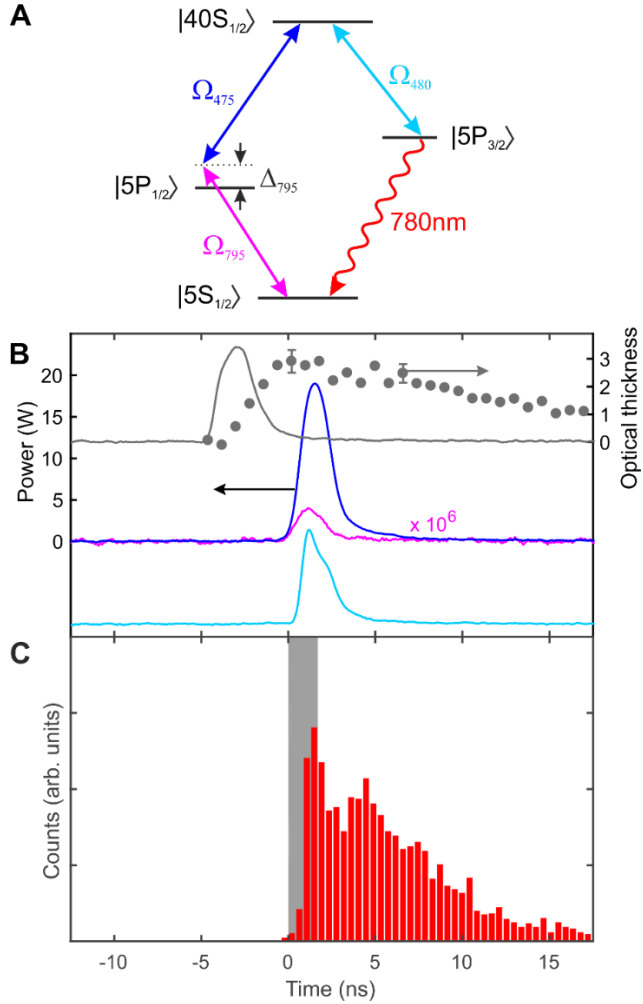

**Fig. 2.** Experiment scheme. **(A)** Excitation level scheme: The excitation path is clockwise starting from $5P_{1/2}$ to $5P_{3/2}$ via $40S_{1/2}$. **(B)** Pulse sequence: the desorption pulse (gray) leads to a drastic increase of the optical depth within 5 ns (gray dots). The optical thickness is determined via statistical measurements at the single-photon level. Excitation pulses (magenta, blue), de-excitation pulse (cyan). Desorption pulse and de-excitation pulse are plotted with an offset relative to the power axis (left). **(C)** Typical signal shape retrieved from a cell length of 1.14 µm. Non-classical emission is observed in an early time interval of the signal, marked in gray color. More details in Supplementary Material.

In figure 2C, a histogram of photoelectric detection events is shown. The signal is mainly composed of two kinds of origins: (I) early photons that are coherently generated in the four-wave mixing process with a decay time of around 1 ns; (II) partly incoherently generated photons that take over at larger delays after the excitation pulses and are mainly originating from collisions of excited atoms with the glass windows. Both processes are much faster than the natural lifetime of the excited atoms. These two kinds of photons do not only differ in their temporal appearance but also in the spatial emission pattern. The reason is that the four-wave mixing photons emit into the excitation mode whereas the background photons are assumed to emit in the full solid angle.

In figure 3A, a normalized second-order intensity correlation function $g^{(2)}(\tau)$ is shown, where the discrete delay is given in multiples of the repetition time (20 ms). The detection events are limited to a time window of 0 to 1.7 ns (blue area in figure 2). All measurements from cell thicknesses 0.94 to 1.14 µm are concatenated. At zero delay, we observe clear anti-bunching with $g^{(2)}(\tau=0) = 0.21(7)$, where the error is the statistical uncertainty given by the error propagation of +/- 1 standard deviation of the correlation fluctuations at $\tau \neq 0$. The four-wave mixing should give $g^{(2)} = 1$, so this is a clear evidence for suppression of multiple excitations of the $40S_{1/2}$ Rydberg states. The detection probability in the measurement shown in figure 3A is on average 0.60%. Accounting for all attenuating elements between generation (behind the first lens) and detection of the photons, like the coupling into a single-mode fiber, imperfect spectral filtering, and finite detection efficiencies of the single-photon counters, we obtain a mean generation efficiency (brightness) of $\varepsilon = 4.0\ \%$.

The effect of the excitation volume can be seen in figure 3B. Here the zero-delay photon-pair correlation is depicted as a function of the cell length L. The blue data points show pair correlations in the time window 0 to 1.7 ns and the red data points show the time span of 1.7 to 10 ns as reference. When the cell length is below a characteristic length $L_0 = 1.16(1)\ \mu m$, Rydberg-Rydberg interaction will efficiently prevent more than one Rydberg excitation in the whole excitation volume. This value $L_0$ is related but not equivalent to the blockade radius. As a comparison to the strong interactions, we also investigate the $32S_{1/2}$ Rydberg state with $r_B = 0.85(13)\ \mu m$. We obtain Poissonian light statistics for all cell lengths around 1 µm and above.

The evolution of the Rydberg blockade with time can be seen in figure 3C. It shows the photon-pair correlation versus the upper limit of the correlation time window that has a constant duration of 1.7 ns. At cell lengths smaller than $L_0$ (blue), $g^{(2)}(\tau=0)$ reveals anti-bunching again. The coherent Rydberg blockaded many-body state lasts about 1.9 ns, then the imprint of the anti-bunching vanishes. As a comparison with negligible Rydberg blockade, the photon-pair correlation at larger cell lengths $L > L_0$ is shown in red.

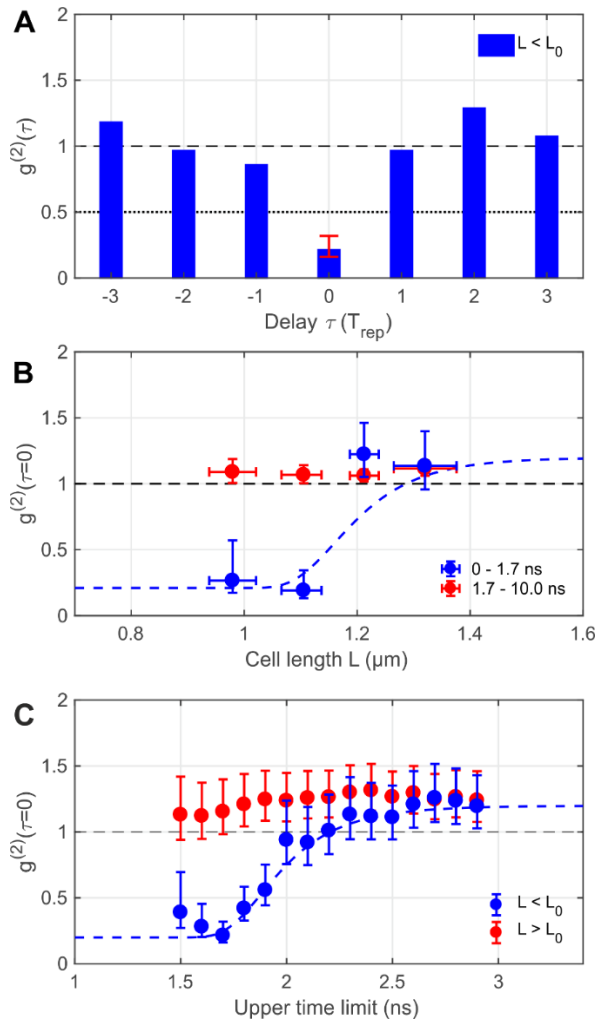

**Fig. 3.** Normalized second-order photon correlation functions. **(A)** As function of coincidence delay time in multiples of the cycle repetition time 20 ms, time window 0 to 1.7 ns, cell thickness $L < L_0$, $g^{(2)}(0) = 0.21(7)$. **(B)** Dependance on the cell thickness in time window 0 to 1.7 ns (blue) and 1.7 to 10 ns (red) as a reference. The blue dashed line is a guide to the eye based on a generic model $g^{(2)}(0) = [1-g^{(2)}_{bg}]\exp[-L_0/L)^\alpha] + g^{(2)}_{bg}$ *(22)*, which is related to the hard-sphere character of the blockade. The characteristic length is $L_0 = 1.16\ \mu m$. **(C)** Dependance on the upper limit of the time window, cell thicknesses $L < L_0$ (blue) and $L > L_0$ (red) as a reference. The blue dashed line is a guide to the eye and is described by a generic model similar to that of figure 3B. Error bars in y-direction obtained by error propagation of +/- 1 standard deviation due to statistical fluctuations; error bars in x-direction mark the span of x-positions of the cell setting the individual data points.

We have shown that strong Rydberg interactions in a hot atomic vapor cell can be used to generate single photons on demand. Advantages of that approach include: (I) no necessity for laser cooling and integrability; (II) effective emission in the forward direction and the potential for storable and identical photons. As next steps, we will therefore determine the indistinguishability of the generated photons, apply optical integration technologies, and combine our source to photon memories based on atomic vapors.

**Acknowledgments:** We thank Y.-H. Chen for her contributions in the early stage of the experiment, F. Schreiber for crafting the vapour cell, N. Gruhler for the coating of the vapour cell with sapphire, C. Müller and M. Gryzlova for contributions to figure 1, P. Michler and M. Schwartz for useful discussions about single-photon sources, as well as H. Alaeian for reading and commenting the manuscript. **Funding:** This work was funded by the ERC under contract number 267100 and the BMBF within Q.com-Q (project number 16KIS0129). **Author contributions:** The experiment was conceived by all authors; execution of the experiment as well as data analysis were performed by F.R.; F.R. wrote the manuscript with contributions from all authors. **Competing interests:** The authors declare no competing financial interests. **Data and materials availability:** All data is available in the main text or the supplementary materials

# Methods

Tight focus
The aspheric lens that confines the 795 nm-excitation volume has a numerical aperture NA = 0.55, an effective focal length $f_{eff}$ = 3.89 mm and is diffraction limited in the focal spot. It is adapted to a laser window thickness of 1.20 mm, which is in our case the window of the vapor cell. The working distance, i.e. the distance between lens surface and focus, is $d_w$ = 2.714 mm. The 795 nm beam diameter results then in $w_0$ = 1.45(5) µm, which is measured by means of a 3D-Piezo stage and an 0.5 µm pinhole.
The collimation lens is of the same type as the focusing lens. The position of this lens is very crucial, so it is placed on a 3D-Piezo stage. Its position is aligned by optimizing to the coupling of the collimated excitation beam into the single-mode fiber that is then guided to the detection setup. The alignment procedure is repeated every 15 to 60 minutes.

Vapor cell
The outer dimension of the cell is 25 mm x 50 mm. The cell windows are made of fused silica. The thickness of the glass plates is d = 1.21(1) mm so that it is matched to the laser window thickness of the aspheric high numeric aperture lenses. The glass plates are coated by a sapphire protection layer with a thickness of 14 nm. This coating protects the cell walls against photo-chemical reactions of fused silica with rubidium mediated by the intense pulses of the four-wave mixing cycle. The coating process has been done by N. Gruhler (at the Institute of Nanotechnology, Karlsruhe Institute of Technology, Germany; currently at the Institute of Physics, University of Münster, Germany) via atomic layer deposition.
The cells are handcrafted by melting the glass plates together. Liquid rubidium is stored in a reservoir tube which is connected to the cell. The whole cell system is at a pressure of about $10^{-8}$ mbar. The temperature of the main part of the cell is set to 160 °C during the measurements and the reservoir is at 140 °C.

The cell has to be positioned precisely relative to the focus point, as the Rayleigh length is only $z_R = 2.3$ µm. This is done via scanning the cell position in the direction of the optical axis and picking the position of strongest saturation in the $D_2$ absorption spectrum. The cell is moved by means of a stepper motor with a precision of about 0.2 µm.

Since the enhanced steady-state density of rubidium is crucial for positioning the cell relative to the focus via saturation spectroscopy, this is the only reason why the cell is heated above room temperature. It is not crucial for the actual four-wave mixing process. For that the atomic density is reached optically via light-induced atomic desorption (LIAD).

Determination and choice of the cell length
The cell is designed in a wedge-shaped geometry providing a range of distances of the glass windows from touching each other up to about 100 µm. Due to the small but finite reflectivity of the glass windows, etalon fringes are observed in transmission, when the cell is moved perpendicular to the optical axis. The finesse of this Fabry-Perot interferometer is $F \approx 0.6$. The cell length is determined by combining the information obtained from the amplitude of the $D_2$ saturation spectroscopy profile and from the etalon fringes of the cell.

The cell is moved in the two perpendicular directions to the optical axis by means of two stepper motors with a precision of about 0.2 µm. In the region of the cell used for the measurements, the thickness changes about 0.5 µm per mm movement. The gradient along the beam diameter can be neglected.

Four-wave mixing cycle
The actual four-wave mixing cycle (figure 2A and B) starts about 5 ns after the desorption pulse. The two-photon transition from ground state $5P_{1/2}(F=3)$ to Rydberg state $40S_{1/2}$ is driven by two 2.5 ns pulses at 795 nm and 475 nm. They are frequency-detuned by $\Delta_{795}/2\pi = 1.36$ GHz to the intermediate state, and the Rabi frequencies are $\Omega_{795}/2\pi = 1.29(17)$ GHz and $\Omega_{475}/2\pi = 0.87(19)$ GHz. In the limit of adiabatic elimination of the intermediate level ($\Delta_{795} \gg \Omega_{795}, \Omega_{475}$), the effective two-photon Rabi frequency is $\Omega_{eff}/2\pi = 0.41$ (16) GHz. The excitation of the Rydberg level is then coupled to the state $5P_{3/2}$ by a retrieval pulse (480 nm, 2.5 ns) with a delay of 0.25(5) ns on average and a Rabi frequency $\Omega_{480}/2\pi = 1.15(9)$ GHz. Both blue pulses have a beam waist of 16(1) µm and are thus much larger than the excitation pulse at 795 nm. The photons at 780 nm are emitted into the same spatial mode as the 795 nm beam.

Excitation pulses
The 480 nm and 475 nm pulses are generated with two dye amplifier systems. Both dye amplifiers are optically pumped by ns-pulses at 355 nm irradiated from a Continuum Powerlite II Nd:YAG laser running at a repetition rate of 50 Hz. The dye amplifier cells are each seeded continuous-wave by two narrow-band laser systems (Toptica TA-SHG) at 480 nm and 475 nm respectively.

The dye amplifier for 475 nm is a Bethune cell which is passed twice by the seed beam *(37)*. The generated pulses are spatially filtered by diamond pinholes and coupled into a single-mode fiber so that the level of amplified spontaneous emission is below 2%. The dye amplifier for 480 nm is a commercial multi-stage amplifier from Sirah. In addition, the pulses are reflected from a spontaneous Brillouin scattering cell before the last amplification step. As a consequence, they get spatially mode-filtered and the amplified spontaneous emission background is removed. The frequency shift caused by the scattering process is considered in the frequency locking scheme.

The pulse energies are between 10 nJ and 45 nJ at pulse durations of 2.5 ns. The pulses temporally jitter up to +/- 1.5 ns. For light-induced atomic desorption, a pulse at 480 nm with about 20 nJ is applied. The timing difference between 475 nm and 480 nm is set by a variable optical delay line *(38)*.

The 795 nm pulses are generated with a free-space Pockels cell, which amplitude-modulates the cw laser beam delivered from a Toptica diode laser systems. The pulse energy is 10 fJ at a duration of 2.5 ns and the pulses jitters less than 0.1 ns.

Data evaluation

We post-select the cycle events with respect to the pulses that are generated in the self-built dye amplifier for a technical reason. The fast part of the 475 nm-pulse jitter is restricted to a range of 0.66 ns. Moreover, the pulse intensities are restricted to a range of about 11 to 17 W/cm² (475 nm pulse) and 4.5 to 9.5 W/cm² (desorption pulse, proportional to 480 nm pulse).

The $g^{(2)}$ function is obtained by restricting the counts to the given time window and digitizing them, which means 1 if there is a counting event and 0 if not. With this the cross-correlation function between the two single-photon counter modules is calculated. The values for the normalized $g^{(2)}(\tau=0)$ function are converted into a histogram to which a Gaussian function is fitted. The standard deviation of this is taken for the error propagation of the $g^{(2)}(\tau=0)$ value.

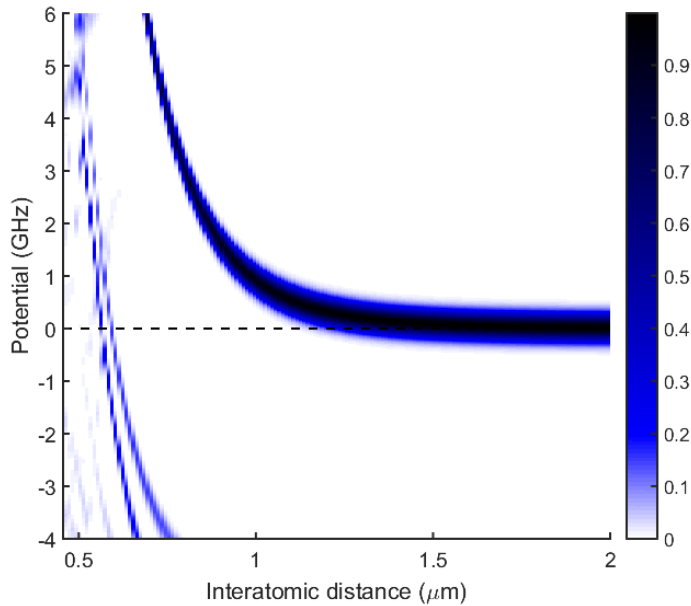

**Fig. S1.** Pair state potential around the 40S-40S resonance of rubidium. 40S-40S admixture versus interatomic distance and potential energy (32). The lower cut-off of the distance corresponds to the Le Roy radius. The Gaussian excitation bandwidth is 400 MHz.

# Supplementary Text

Rydberg blockade

The Rydberg blockade is achieved by the energy shift visible in the pair state potentials of the Rydberg states *(32)*. The potentials are calculated via diagonalization of the interaction Hamiltonians including many neighboring states. In figure S1 the 40S-40S pair state potential in dependence on the interatomic distance is depicted. The important quantity is the 40S-40S character of pair states. For larger distances, only the 40S-40S state is relevant. It shows a van der Waals-type scaling $\sim C_6/r^6$. A fit to the numerical calculations results in $C_6 = 2\pi\hbar \times 806.3$ MHz µm$^6$. For smaller distances (around 0.6 µm) other pair states ($39P_{1/2,3/2}$-$40P_{1/2,3/2}$) with 40S-40S admixtures move in, however only with a small admixture amplitude. The calculations are valid above the Le Roy radius, where the electron wave functions of the individual atoms start to overlap. For the 40S state of rubidium it is $R_{LR} = 0.46$ µm *(32)*.

Four-wave mixing signal
The measured statistical signals of the individual data points of figure 3B are shown in figure S2. The signal is composed of a four-wave mixing part in the beginning and a background part at later times (we have strong hints that this background is a consequence of spin-flips due to atom-wall collisions). The photon number of the four-wave mixing signal is proportional to the square of the cell length. Note that the four-wave mixing signal is rather an induced than a spontaneous process. This enables constructive interference between the many individual atomic emitters in the beginning of the signal. The temporal decay of the signal is then originated in the destructive interference and thus much faster than the natural decay time of the excited state of rubidium, which is 26.2 ns *(39)*.

The peak timing of the collisional part changes with cell length. It is assumed that collisions of $5P_{1/2}$-excited atoms with the glass walls lead to spin flips and thus to emission from the $5P_{3/2}$ state. For shorter cell lengths the four-wave mixing signal and the collisional background start to overlap. This limits the fidelity of the single-photon source, as only the four-wave mixing signal is anti-bunched.

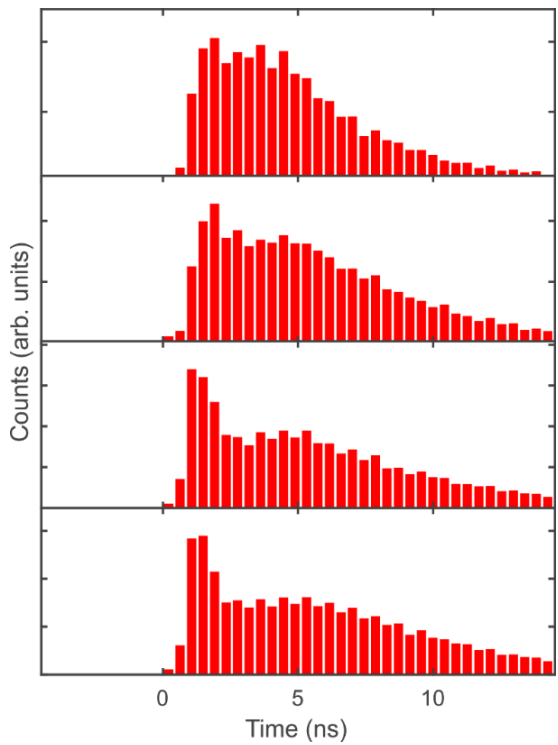

**Fig. S2.** Shapes of statistical four-wave mixing signal counts for the cell lengths of figure 3B; increasing cell length from top to bottom.

Limitations of anti-bunching
In this proof-of-concept experiment, we see several limitations for a pure anti-bunching, which will be overcome in the future. Due to the temporal overlap of excitation and de-excitation pulses, a small population fraction is directly transferred to the final state without populating the Rydberg state and therefore without excitation blockade (Raman transition). This can be improved with optimal pulse shapes and timings *(34)*. Additionally, residual double-excitations can occur, as the excitation volume and the blockade volume are of similar size. With higher-lying Rydberg states, the blockade volume can be increased.

Limitations of brightness
The brightness of the source is limited by the finite excitation efficiency (in our case about 30%). This can be enhanced by optimized intensities, shapes, and timings of the excitation pulses *(34)*. Further, the temporal and spatial mode overlap of signal and background (figure 2C) forces us to truncate the retrieved four-wave signal in time. More sophisticated desorption schemes could reduce or shift the collisional background to later times. Finite size effects (volume size compared to the optical wavelength, finite number of atoms) lead to imperfections in the emission pattern and hence limit the directivity. Simulations show that the emission fraction into the collection mode is above 90 %. The finite size and thus microscopic fluctuations of numbers, positions, and velocities of the atoms might also limit the indistinguishability between separate single photons. All these issues will be the subject of further studies.

Limitations due to laser source
The major technical bottleneck of our approach is the laser system that has been used to excite the atoms to their Rydberg states. It provides only a repetition rate of 50 Hz, accompanied by timing jitters that require an expensive post-selection process. Therefore, in the next generation of this experiment, we will modify the excitation scheme such that we can apply well-developed high-power near-infrared laser systems to the Rydberg transition. With this, we will reach repetition rates up to 10 MHz and theoretical calculations predict generation efficiencies close to 100 %.